\documentclass[preprint,showpacs]{revtex4-1}
\usepackage{amsmath}
\usepackage{amsfonts}
\usepackage{graphicx}
\usepackage{caption}

\begin{document}

\title{Effects of the continuous spontaneous localization model in the regime of large localization length scale}
\date{\today}
\author{D.~J.~Bedingham}
\email{d.bedingham@imperial.ac.uk}
\affiliation{Blackett Laboratory \\ Imperial College \\ London SW7 2BZ \\ United Kingdom}
\pacs{03.65.Nk, 03.65.Ta, 03.75.-b, 05.10.Gg}

\begin{abstract}
Working in the limit in which the localization length scale is large compared to other relevant length scales we examine three experimental situations with the continuous spontaneous localization (CSL) model---a well-motivated alternative to standard quantum theory. These are the two-slit experiment; scattering from a potential barrier; and the release of two non-interacting particles simultaneously from a potential trap. In each case we calculate the diagonal part of the time evolved density matrix giving a probability density function over final measured states. The case of the two-slit experiment is already well understood and we reproduce some known conditions for observing loss of interference. The other examples have not previously been examined in the context of CSL. For scattering from a potential barrier we find that the probability of reflection is unchanged by CSL, however, the momentum state is spread in a characteristic way. For the case of two particles released simultaneously from a trap we find that it is more likely that the particles diffuse in the same direction than would happen if the particles behaved independently. We assess the possibility of observing these effects.

\end{abstract}
\maketitle

%%%%%%%%%%%%%

\section{Introduction}

Spontaneous localization (SL) is an alternative to standard quantum theory motivated by the measurement problem \cite{GRW,CSL1,CSL2} (for reviews see \cite{REP1,REP2}). In SL the quantum state has the status of a real physical object. The state evolves according to a stochastic generalization of the Schr\"odinger equation and behaves in such a way that certain macro superposition states are unstable. The theory itself determines precisely which types of state are unstable and remarkably gives a universal account of phenomena which so far agrees with our experience ranging from the behaviour of quantum particles to the classical macroscopic world.

Spontaneous localization is particularly interesting because it makes several predictions which are in conflict with standard quantum theory. This means that there is the possibility to experimentally test it. For example, SL predicts the collapse of a superposition of quasi-localized wave packets for a particle (or composite object) of sufficiently large mass. The way to observe this process would be to look for a characteristic loss of interference when the packets are brought together. These kinds of experiments are being developed as part of a general program to observe quantum interference for objects of increasing mass. One promising experiment involves diffraction of a beam of Au clusters by a laser grating \cite{ARNDT}; another involves attempting to put a small mechanical device in a superposition of different position states \cite{MIRR} (see also \cite{KIP}). 

In this article we focus on the continuous spontaneous localization (CSL) model \cite{CSL1,CSL2} and examine a selection of experimental scenarios in the limit is which the localization length scale is large compared to other length scales (such as the spatial extent of the wave function). The reason for taking this limit is primarily to make calculations easier to perform. This enables us to obtain precise results within the regime of validity. However, we note that at present there is no upper bound on the CSL length scale imposed by experimental results \cite{TUM2}. Our considerations might therefore be seen a way to further constrain the parameter space of the CSL model. 

In the CSL model the state process is expressed as a stochastic differential equation for the state vector (see Sec.~\ref{S1}). Stochastic changes in the state are governed by a classical noise process and for a given realization of this noise process the state vector follows a unique trajectory. However, it is convenient to consider a statistical ensemble of systems describing the result of running the same experiment with the same initial conditions many times. The dynamics then turn an initial pure state into a mixture of states, each resulting from a different realized noise. The appropriate way to describe this is in terms of a density operator 
\begin{align}
\hat{\rho}_t = \mathbb{E}[|\psi_t\rangle\langle \psi_t |],
\label{dop}
\end{align}
where $\mathbb{E}$ denotes classical expectation over states. 

The density matrix can be written in terms of a density matrix propagator which for a single particle system is expressed in coordinate space as
\begin{align}
\rho_t(x,y) = \int dx'dy' J(x,y,t|x',y',t')\rho_{t'}(x',y'),
\label{rhosol1}
\end{align}
where $\rho_t(x,y) = \langle x|\hat{\rho}_t|y\rangle$ and $t>t'$. The advantage of working with the density matrix propagator is that we can separate out the dependence on the initial condition. There are well established methods to determine the density matrix propagator, see e.g.~\cite{CALD,ANAS}. Here we will demonstrate that in the case where the system is more localized than the CSL length scale, the CSL master equation can be reduced to a form where the propagator is a Gaussian function. This implies that for situations in which the initial wavefunction is of Gaussian form (or a superposition of Gaussians), the final density matrix state can be straightforwardly determined by performing Gaussian integrals over the variables $x'$ and $y'$. 

After an introduction to the CSL model in Sec.~\ref{S1} we apply the density matrix propagator technique to the two-slit experiment in Sec.~\ref{Stwoslit}. We consider an initial wavefunction composed of two separated Gaussian peaks corresponding to the particle passing through each of the two slits. We then allow the wave packets to spread and overlap and determine the probability distribution for the location of the particle. We find that the expected interference pattern degrades as a result of CSL as either the mass of the particle increases, the slit width increases, or the separation between slits increases. We highlight the result that a cubic improvement on the degradation effect can be made by increasing the slit separation provided it is less than the CSL length scale. 

The same techniques also allow us to consider interactions with more complicated potentials by treating the potential perturbatively. We demonstrate this in detail in Sec.~\ref{SScat} where we consider a particle undergoing CSL dynamics as it is scattered from a small potential barrier. We calculate the effect of the potential up to second order. This turns out to be the lowest order necessary in order to see reflection from the barrier. Reflection from a barrier whose height is small compared to the energy of the incoming particle is a quantum mechanical effect which we might expect to be diminished by CSL. However, we find that there is no appreciable reduction in the probability for reflection when compared to a standard quantum calculation. The significant difference is that the wave packet undergoes a characteristic spreading in momentum space. This is expected to be small given standard estimates for the CSL parameters and the scattering does nothing to enhance the order of magnitude of the momentum diffusion over that which would occur for a free packet without the potential. 

In Sec.~\ref{S2} we apply the density matrix propagator technique to a situation involving $2$ particles. In CSL the localization mechanism acts on the total particle number density state rather than on each particle individually. Thus implies that for a system of 2 non-interacting particles, the localization mechanism will lead to correlations in the diffusion undergone by each particle separately. In order to demonstrate this we present the $2$-particle propagator and solve for a situation in which the initial wavefunctions of the particles perfectly overlap. We find that the final distribution of particle locations shows that the particles are more likely to be found closer together than would be expected if the particles behaved independently. We examine the possibility of this being used as an new experimental test of CSL and consider the expected scales of magnitude required to see the effect. We end with some discussion and summary in Sec.~\ref{SDISC}.

%%%%%%%%%%%%%

\section{Continuous spontaneous localization}
\label{S1}
Here we present the CSL model in terms of a diffusion process for the quantum state and show how it can be represented in terms of a deterministic master equation for the stochastically averaged density matrix (\ref{dop}). The CSL model is a non-relativistic model involving quantum fields. We will write down the master equation for the case of a one particle excitation and present the density matrix propagator. Later in Sec.~\ref{S2} we will do the same for a two identical particle state. For simplicity we work in 1 dimension. The results should be valid for the 3-dimensional  theory since the different dimensions decouple in the equations of motion. 

We introduce two parameters $\lambda$ and $1/\sqrt{\alpha}$. These are, respectively, the rate and the length scale of spontaneous localization. The state evolution is then described by the quantum state diffusion equation \cite{CSL1,CSL2}
\begin{align}
d|\psi_t\rangle =& \left[-\frac{i}{\hbar} \hat{H}- \frac{\lambda}{2}\int d{x} \left(\hat{N}({x}) - \langle \hat{N}({x}) \rangle\right)^2\right]dt |\psi_t\rangle
\nonumber\\
&+ \sqrt{\lambda}\int d{x}\left(\hat{N}({x}) - \langle \hat{N}({x}) \rangle\right) dB_t ({x})|\psi_t\rangle,
\label{CSL}
\end{align}
where the number density operator $\hat{N}({x})$ is given by
\begin{align}
\hat{N}({x}) = \left( \frac{\alpha}{\pi}\right)^{1/4}\int d{y}\exp\left\{-\frac{\alpha }{2}({x}-{y})^2 \right\}\hat{a}^{\dagger}({y})\hat{a}({y}),
\label{N}
\end{align}
the field annihilation and creation operators $\hat{a}({x})$ and $\hat{a}^{\dagger}({x})$ satisfy
\begin{align}
[\hat{a}({x}),\hat{a}^{\dagger}({y})] = \delta({x}-{y}),
\end{align}
and the field of Brownian motions satisfy
\begin{align}
\mathbb{E}[dB_t({x})] = 0;\quad dB_s({x})dB_{t}({y}) = \delta_{st}\delta({x}-{y})dt.
\end{align}
Equation (\ref{CSL}) can be thought of as a modification of the Schr\"odinger equation in order to describe both unitary and state reduction-type behaviour. The parameter $\alpha$ is taken to be a constant. Ghirardi, Rimini and Weber (GRW) give an estimate of $1/\sqrt{\alpha} = 10^{-7}{\rm m}$ \cite{GRW}. Arguments related to experimental bounds on the spontaneous emission of photons from Germanium suggest that $\lambda\propto m^2$ \cite{M2}, where $m$ is the particle mass. We therefore assume that 
\begin{align}
\lambda = \left(\frac{m}{m_0}\right)^2\lambda_0,
\label{lamb}
\end{align}
where $m_0$ is the nucleon mass and $\lambda_0$ is the rate of spontaneous localization for a nucleon. By assuming (\ref{lamb}) we are acknowledging that the number density operator in fact represents the mass density. The estimate of GRW is $\lambda_0 = 10^{-16}{\rm m}$. 

We note that the GRW estimates are not definitive. A study of the range of parameter space compatible with experiments can be found in Ref.~\cite{TUM2}. In particular $1/\sqrt{\alpha}$ may be much larger than $10^{-7}{\rm m}$. This is of relevance since we will work in an approximation where the particle is more localized (i.e.~its wavefunction is narrower) than the CSL length scale.

The density operator is given by Eq.(\ref{dop}). It follows from (\ref{CSL}) that the density operator satisfies the master equation
\begin{align}
\frac{\partial\hat{\rho}_t}{\partial t} = -\frac{i}{\hbar}\left[\hat{H},\hat{\rho}_t \right] 
- \frac{\lambda}{2}\int dx\left[\hat{N}(x),\left[ \hat{N}(x),\hat{\rho}_t \right]\right].
\label{Mop}
\end{align}

Consider the case where the state is composed of 1 particle. A single particle state has the form
\begin{align}
|\psi\rangle = \int d{x}\; \psi({x}) \hat{a}^{\dagger}({x})|0\rangle,
\end{align}
where $|0\rangle$ is the vacuum state. Here we can identify $\psi({x})$ as the wave function for the particle and improper position eigenstates take the form $|{x}\rangle = \hat{a}^{\dagger}({x})|0\rangle$. The Hamiltonian is given in position space as
\begin{align}
H = -\frac{\hbar^2}{2m} \frac{\partial^2}{\partial x^2} + V(x).
\label{Hx}
\end{align}
Then using (\ref{Mop}) with (\ref{N}) and (\ref{Hx}), the master equation can be written as
\begin{align}
\frac{\partial}{\partial t}\rho_t(x,y) =& \frac{i\hbar}{2m}\left(\frac{\partial^2}{\partial x^2} - \frac{\partial^2}{\partial y^2}\right)\rho_t(x,y)
-\frac{i}{\hbar}\left(V(x) - V(y) \right)\rho_t(x,y)
\nonumber\\
&-\lambda \left(1-\exp\left\{-\frac{\alpha }{4}({x}-{y})^2\right\}\right)\rho_t(x,y),
\end{align}
where $\rho_t(x,y)=\langle x |\hat{\rho}| y \rangle$. 

Next we assume that $x-y$ is typically much smaller than $1/\sqrt{\alpha}$ so that we can simplify to 
\begin{align}
\frac{\partial}{\partial t}\rho_t(x,y) =& \frac{i\hbar}{2m}\left(\frac{\partial^2}{\partial x^2} - \frac{\partial^2}{\partial y^2}\right)\rho_t(x,y)
-\frac{i}{\hbar}\left(V(x) - V(y) \right)\rho_t(x,y) \nonumber\\
&-\frac{D}{\hbar^2}({x}-{y})^2\rho_t(x,y),
\label{M1}
\end{align}
where
\begin{align}
D = \frac{\lambda \alpha\hbar^2}{4}.
\label{D}
\end{align}
The advantage of this simplified form for the master equation is that we can solve it exactly in the case where $V = 0$. The solution can be written in terms of the density matrix propagator as in Eq.(\ref{rhosol1}). The result for Eq.(\ref{M1}) is (see e.g. Ref.~\cite{ZOUP})
\begin{align}
J(x,y,t|x',y',t') =& \frac{m}{2\pi \hbar (t-t')}\exp \left\{ \frac{im}{2\hbar(t-t')} \left[(x-x')^2-(y-y')^2 \right] \right\}
\nonumber \\ & \times \exp \left\{ -\frac{D(t-t')}{3\hbar^2}\left[ (x-y)^2+(x-y)(x'-y')+(x'-y')^2 \right] \right\}.
\label{propx}
\end{align}
It is also useful to express the density matrix propagator in momentum space where
\begin{align}
J(p,q,t|p',q',t') =& \frac{1}{\sqrt{4\pi D(t-t')}}\delta(p-q-p'+q')\nonumber\\
& \times	\exp\left\{ -\frac{i(t-t')}{4m\hbar}\left( p^2-q^2+p'^2-q'^2 \right)\right\}\nonumber\\
& \times \exp\left\{ -\frac{1}{4D(t-t')}(p-p')^2 -\frac{D(t-t')^3}{12m^2\hbar^2} (p-q)^2\right\}.
\label{propp}
\end{align}
Notice that the propagators tend to delta functions as $t\rightarrow 0$ and that the free propagator is recovered upon setting $D=0$. The potential $V$ can be introduced perturbatively in a standard way which we outline for the specific case of scattering from a potential barrier in Sec.~\ref{SScat}.

%%%%%%%%%%%%%

\section{Two-slit experiment }
\label{Stwoslit}
In the case where $V = 0$ the master equation is solved exactly with result given by Eqs (\ref{rhosol1}) and (\ref{propx}). We consider a simple two-slit experiment and model only the dimension parallel to a line intersecting the two slits perpendicularly. We approximate the initial wavefunction by two Gaussian peaks each of width $\sigma$ (the slit width) and separated by a distance $2\mu$ (the slit separation). Ideally $\mu \gg \sigma$, such that the two peaks do not overlap. The initial density matrix at time $0$ can then be written 
\begin{align}
\rho_0(x,y) = \psi(x)\psi^{*}(y),
\end{align}
with
\begin{align}
\psi(x) = \frac{1}{\sqrt{2}}\frac{1}{(2\pi\sigma^2)^{1/4}}\left[\exp\left\{-\frac{1}{4\sigma^2}(x-\mu)^2\right\}
+\exp\left\{-\frac{1}{4\sigma^2}(x+\mu)^2\right\}\right].
\end{align}
Note that this choice of initial condition neglects any wavefunction collapse between the source of the particle beam and arrival at the two slit screen. This can be justified by assuming that the beam is sourced by a slit much narrower than $\sigma$ which spreads the wavefunction much more rapidly so that the particle passes through the preceding section of the interferometer in a comparatively negligible time.

We suppose that the particle reaches a measuring screen beyond the two slits after travelling for a time $t$ (in a spatial dimension that we do not consider here). The probability density for finding the particle at position $x$ at this time is given by $\rho_t(x,x)$. After setting $x = y$ the result of the integrals in Eq.(\ref{rhosol1}) is
\begin{align}
\rho_t(x,x) =& \frac{1}{2}\frac{1}{(2\pi K\sigma^2)^{1/2}}\nonumber\\
& \times \left[ \exp\left\{-\frac{1}{2K\sigma^2}(x-\mu)^2\right\}+\exp\left\{-\frac{1}{2K\sigma^2}(x+\mu)^2\right\}
\right.\nonumber \\
& \left. \quad+2\cos\left\{\frac{\hbar t\mu}{2mK\sigma^4}x\right\}
\exp\left\{-\frac{1}{4K\sigma^2}\left[(x-\mu)^2+(x+\mu)^2+\frac{4Dt^3\mu^2}{3m^2\sigma^2}\right]\right\}
\right],
\label{twoslit}
\end{align}
where
\begin{align}
K = \frac{2Dt^3}{3m^2\sigma^2} + \frac{\hbar^2 t^2}{4m^2\sigma^4} + 1.
\label{K}
\end{align}

\begin{figure}[h]
        \begin{center}
        	\includegraphics[width=15cm]{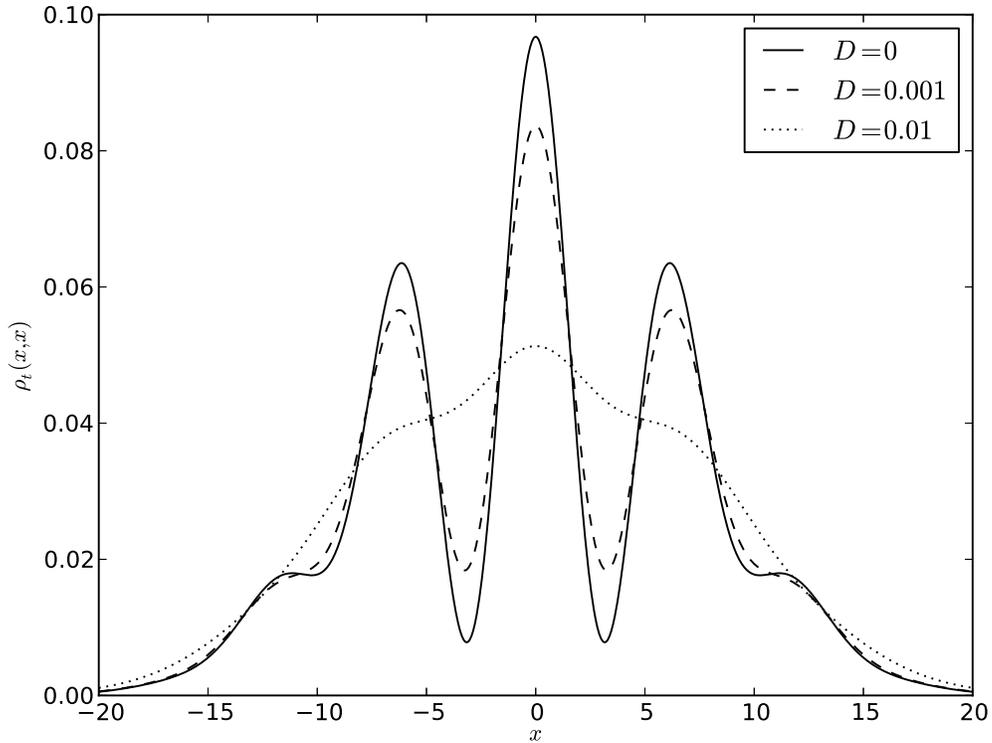}
        \end{center}
\caption{Two-slit probability distribution function for different values of the CSL parameter, $D = \lambda \alpha\hbar^2/4$.}
\label{F2}
\end{figure}

The parameter $K$ gives a measure of the spread of each of the two peaks. The two time-dependent contributions are due respectively to spontaneous localization and standard quantum dispersion. It is reasonable to assume that the dominant contribution is that due to standard quantum dispersion. We can then estimate the time scale on which the two peaks reach a state of significant overlap. This occurs when $\sqrt{K\sigma^2}\sim \mu$ which results in 
\begin{align}
t \sim \frac{m\sigma\mu}{\hbar}.
\label{talb}
\end{align}
If the interference term is suppressed before this time then the interference peaks do not get the chance to develop. From the last term in Eq.(\ref{twoslit}) we see that this happens when 
\begin{align}
D \gtrsim D_{\rm crit} = \frac{\hbar^3}{m\sigma\mu^3}.
\label{Dcrit}
\end{align}

In Fig.~\ref{F2} we display the probability distribution (\ref{twoslit}) for some different values of $D$. We choose units such that $\hbar = m = \sigma = 1$. If we choose $\mu = 5$ then $t = 10$ in these units is of the order of the time taken for the packets to overlap. We have plotted the probability distribution for three values of $D$ at this time. The interference peaks disappear somewhere between $D = 0.001$ and $D = 0.01$. This agrees with the estimate from Eq.(\ref{Dcrit}) of $D_{\rm crit} = 0.008$.

Since from (\ref{lamb}) and (\ref{D}) it is expected that the parameter $D$ increases with mass as $m^2$, there are three things that can be done to improve the chances of observing interference loss: (i) increase the mass $m$; (ii) increase the distance between slits $\mu$; and (iii) increase the slit width $\sigma$. The first two ways offer the greatest improvement since they each have a cubic effect. The other way to eliminate interference effects is to make $\sigma$ larger. This has the effect of causing the wavefunction peaks to spread at a slower rate and therefore effectively buys more time for the system to undergo spontaneous localization. We note that for $\mu > 1/\sqrt{\alpha}$ the form of (\ref{Dcrit}) would be significantly different (e.g.~see Eq.(4) in \cite{ARNDT} or Sec.~VII B in \cite{SIM2} for comparison).

Note that if $D \sim D_{\rm crit}$ then the assumption that the density matrix spreading is dominated by standard quantum dispersion on the time scale  $m\sigma\mu/\hbar$ is confirmed provided that $\mu \gg \sigma$. For $\mu\sim\sigma$ the diffusive spreading due to spontaneous localization is of the same order as the standard quantum dispersion for this value of $D$ on this time scale, however, the argument leading to Eq.(\ref{Dcrit}) will still hold. 

The two-slit experiment is the standard example of how SL differs in its predictions from standard quantum theory. It is also one of the best opportunities to actually test SL. The fact that the interference degradation effect is improved as the mass cubed or the slit separation cubed means that these parameters provide effective levers for entering the SL regime.

A planned experiment aims to use $10^{8}{\rm amu}$ Au clusters and a laser grating of wavelength $157{\rm nm}$ ($\mu=\sigma=78.5{\rm nm}$) \cite{ARNDT}. These values correspond to $D_{\rm crit}/\hbar^2\sim 10^{-2}{\rm m}^{-2}{\rm s}^{-1}\times(m/m_0)^2$. This is equivalent to the order of magnitude of the GRW estimates of the spontaneous localization parameters ($\lambda_0 = 10^{-16}{\rm s}^{-1}$ and $1/\sqrt{\alpha} = 10^{-7}{\rm m}$) in agreement with the estimations of Ref.\cite{ARNDT}.

%%%%%%%%%%%%%

\section{Scattering from a potential barrier}
\label{SScat}
Here we apply the density matrix propagator technique perturbatively in order to analyse the problem of a particle scattering from a small potential barrier of height $V$. For $V$ much less than the energy of the incoming particle we would classically expect the particle to pass right over the barrier. However, quantum mechanics predicts a small amount of reflection. We would like to consider the effect of spontaneous localization on this process. Since quantum mechanical reflection can be regarded as an interference effect in momentum space we might well expect that SL has the effect of diminishing the amount of reflection. However, the answer turns out to be more subtle. (See also Ref.\cite{JJD} where the related problem of scattering in the presence of a thermal environment is considered.)

We solve Eq.(\ref{M1}) by treating the potential term as a small perturbation. Working in momentum space where the propagator is given by Eq.(\ref{propp}), the zeroth order contribution to the density matrix is
\begin{align}
\rho^{(0)}_t(p,q) = \int dp_0 dq_0 J(p,q,t|p_0,q_0,0)\rho_0(p_0,q_0).
\end{align} 
The probability distribution in momentum space at time $t$ is found by setting $p = q$. We choose an initial density matrix of the form 
\begin{align}
\rho_0(p_0,q_0) =& \sqrt{\frac{2}{\pi}}\frac{\sigma}{\hbar}\exp
\left\{ -\frac{\sigma^2}{\hbar^2}\left[(p_0-\bar{p})^2+(q_0-\bar{p})^2\right] - \frac{i}{\hbar}\bar{x}(p_0-q_0) \right\}
\\ \simeq & \frac{\sqrt{2\pi}\hbar}{\sigma}\delta(p_0-\bar{p})\delta(q_0-\bar{p}),
\label{pureapprox}
\end{align}
with $\bar{p}\sigma/\hbar\gg1$, corresponding to a near perfect plane wave with momentum $\bar{p}$. When using the approximation (\ref{pureapprox}) we must use $t=2m\bar{x}/\bar{p}$ with $\bar{x} = \sqrt{(\pi/2)}\sigma$ to determine the approximate time taken for the wave packet to cross the potential barrier. With this initial state the zeroth order contribution to the probability density function is
\begin{align}
\rho^{(0)}_t (p,p) = \frac{1}{\sqrt{4\pi Dt}}\exp \left\{ -\frac{1}{4Dt}(p-\bar{p})^2\right\}.
\label{zeroth}
\end{align}
This describes the diffusive spreading of the pure mode initial state due to SL. The potential does not appear at this level of approximation and there is no reflection. The momentum spread at zeroth order is 
\begin{align}
\Delta p \sim \sqrt{Dt}.
\end{align}
This is the dominant effect for the transmitted packet.

The first order perturbative correction is given by 
\begin{align}
\rho^{(1)}_t(p,q) =& -\frac{1}{\sqrt{2\pi\hbar}}\int_0^t dt' \int\left[\prod_{i=0}^{2}dp_i dq_i\right] J(p,q,t|p_2,q_2,t')\nonumber\\
&\times\frac{i}{\hbar}\left(V(p_2-p_1)\delta(q_2-q_1)-V(q_2-q_1)\delta(p_2-p_1)\right)\nonumber\\
&\times J(p_1,q_1,t'|p_0,q_0,0)\rho_0(p_0,q_0),
\label{rho2}
\end{align}
where the potential in momentum space is given by
\begin{align}
V(p) = \frac{1}{\sqrt{2\pi\hbar}}\int dx\exp\left\{-\frac{i}{\hbar}px\right\}V(x).
\end{align}
The contributions from the two potential terms cancel out when $p=q$ so at first order there is no contribution to the probability density function $\rho_t(p,p)$. 

The second order term is
\begin{align}
\rho^{(2)}_t(p,q) =& \frac{1}{2\pi\hbar}\int_0^t dt' \int_0^{t'}dt''\int\left[\prod_{i=0}^{4}dp_i dq_i\right] J(p,q,t|p_4,q_4,t')\nonumber\\
&\times\frac{i}{\hbar}\left(V(p_4-p_3)\delta(q_4-q_3)-V(q_4-q_3)\delta(p_4-p_3)\right)\nonumber\\
&\times J(p_3,q_3,t'|p_2,q_2,t'')\nonumber\\
&\times\frac{i}{\hbar}\left(V(p_2-p_1)\delta(q_2-q_1)-V(q_2-q_1)\delta(p_2-p_1)\right)\nonumber\\
&\times J(p_1,q_2,t''|p_0,q_0,0)\rho_0(p_0,q_0),
\label{rho2}
\end{align}
In order to perform the calculation we use a Gaussian potential barrier of width $a$ where
\begin{align}
V(x) = V_0\frac{1}{\sqrt{2\pi a^2}}\exp\left\{-\frac{1}{2a^2}x^2\right\}.
\end{align}
Working to the same order in perturbation theory, standard quantum mechanics predicts that the reflected component from this potential is
\begin{align}
|\psi^{\rm ref}_{\infty}(p)|^2  = \frac{V_0^2 m^2}{\hbar^2 p^2} 
\exp\left\{-\frac{4a^2}{\hbar^2}p^2\right\} |\psi_{-\infty}(-p)|^2.
\label{unit}
\end{align}
For an initial pure state of momentum $\bar{p}$ the reflected state has momentum $-\bar{p}$.

The Gaussian momentum integrals in (\ref{rho2}) can be evaluated and the result after a long calculation is
\begin{align}
\rho^{(2)}_t(p,p) = A_t(p,p) +\left[{A}_t(p,p)\right]^*+B_t(p,p) +\left[{B}_t(p,p)\right]^*,
\label{second}
\end{align}
where
\begin{align}
A_t(p,p) =& -\frac{mV_0^2}{2\pi\hbar^3 \bar{p}}\frac{1}{t}\int_0^t dt'\int_0^{t'} dt'' \frac{1}{\sqrt{K_A}}\nonumber\\
&\times \exp \left\{ -\frac{a^2}{K_A\hbar^2}(p-\bar{p})^2 +\frac{i(t'-t'')}{2K_A m \hbar }(p-\bar{p})^2 \right\}\nonumber\\
&\times \exp \left\{ -\frac{D(t'-t'')^2(t'+2t'')}{3K_A m^2\hbar^2}p^2 - \frac{D(t'-t'')^3}{3K_A m^2\hbar^2}\bar{p}p \right\} \nonumber\\
&\times \exp \left\{ -\frac{D(t'-t'')^2(3t-2t'-t'')}{3K_A m^2\hbar^2}\bar{p}^2 \right\} ,
\label{EA}
\end{align}
and
\begin{align}
B_t(p,p) =& \frac{mV_0^2}{2\pi\hbar^3 \bar{p}}\frac{1}{t}\int_0^t dt'\int_0^{t'} dt'' \frac{1}{\sqrt{K_B}}\nonumber\\
&\times \exp \left\{ -\frac{a^2}{K_B\hbar^2}(p-\bar{p})^2 + \frac{i(t'-t'')}{2K_B m \hbar }(p^2-\bar{p}^2)\right\}\nonumber\\
&\times \exp \left\{ -\frac{D(t'-t'')^2(t'+2t'')}{3K_B m^2\hbar^2}p^2 - \frac{D(t'-t'')^3}{3K_B m^2\hbar^2}\bar{p}p \right\} \nonumber\\
&\times \exp \left\{ -\frac{D(t'-t'')^2(3t-2t'-t'')}{3K_B m^2\hbar^2}\bar{p}^2 \right\} ,
\label{EB}
\end{align}
with
\begin{align}
K_A = \frac{4Da^2}{\hbar^2}t + \frac{D^2}{3m^2\hbar^2}(t'-t'')^2\left[4(t'+2t'')t-3(t'+t'')^2\right]-\frac{2iD}{m\hbar}(t'-t'')t,
\label{KA}
\end{align}
and
\begin{align}
K_B = 1+K_A+\frac{2iD}{m\hbar}(t'-t'')(t'+t'').
\label{KB}
\end{align}

\begin{figure}[h]
        \begin{center}
        	\includegraphics[width=15cm]{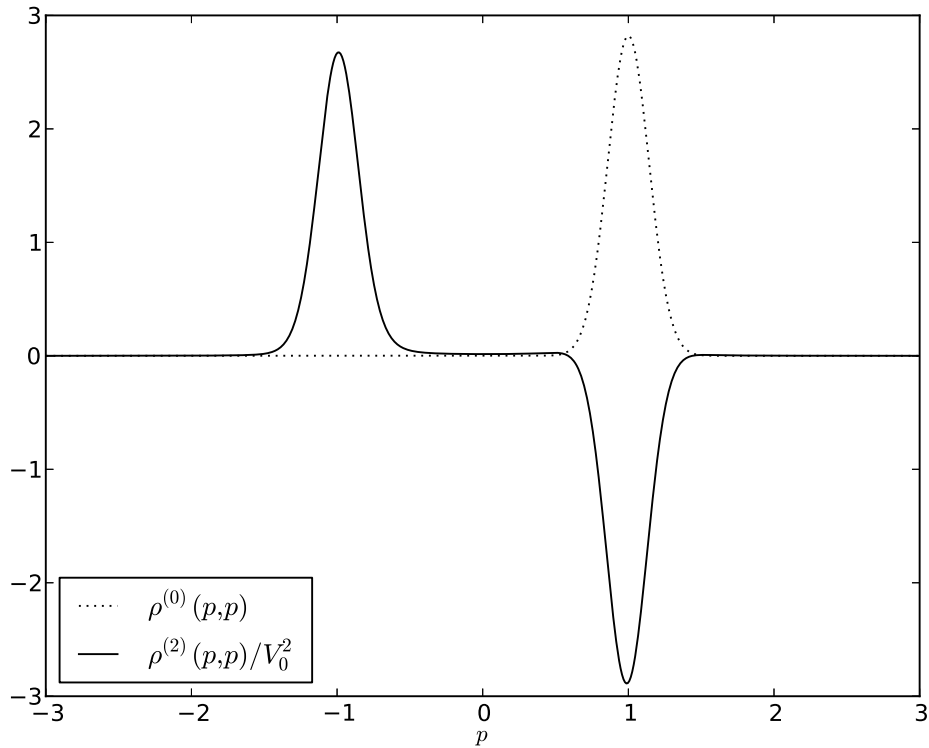}
        \end{center}
\caption{Probability distribution function in momentum for scattering from a Gaussian potential barrier at zeroth order (dashed line) and second order (solid line) in $V$.}
\label{F1}
\end{figure}
The reflected contribution is seen to result from $B$ and $B^*$ and agrees with Eq.(\ref{unit}) as $D\rightarrow 0$. The time integrals can be performed numerically and an example is shown in Fig.~\ref{F1}. For the plots we choose units such that $\hbar=m=\bar{p} = 1$ and set other parameters to be $t=100$, $D = 0.0001$, and $a = 0.1$. The initial state corresponds to a delta function peak at $p=1$. After the scattering has taken place there are two peaks corresponding to transmission and reflection. The graph show the zeroth order contribution (pure transmission) and the second order contribution (scaled by the potential height in order to compare the contributions at the same scale). There is a close but not exact agreement between the zeroth order transmitted peak and the negative of the transmitted peak at second order. We find that the total contribution at second order integrates over momentum to 0 as expected.

Let us now examine the reflected peak. From Eqs (\ref{EB}), (\ref{KA}) and (\ref{KB}) we can read off the time scales
\begin{align}
t_E &= \frac{m\hbar}{\bar{p}^2}, \\
t_1 &= \frac{m\hbar}{Dt}, \\
t_2 &= \frac{m\hbar}{\bar{p}\sqrt{Dt}},
\end{align}
where $t$ is the total time taken for the packet to cross the barrier and for now we ignore $a$. The time scale $t_E$ is the energy time. It can be understood as the time for the packet to traverse its own wavelength. This time is of most relevance to the formation of the reflected packet and if the effects of CSL are to prevent the formation of the reflection they must act faster that $t_E$ \cite{JJD}. However, if we impose the condition
\begin{align}
\bar{p} \gg \Delta p,
\label{momcon}
\end{align}
essentially saying that momentum diffusion of the free packet over time $t$ does not effect its integrity as the wavepacket, then we find that
\begin{align}
t_1 \gg t_2 \gg t_E.
\end{align}
The unitary part of the propagator contributes to the time integral in (\ref{EB}) on a time scale $t_E$ whereas the CSL terms only contribute at times $\sim t_2\gg t_E$. This shows that in the regime of momentum fluctuations small enough that the results make physical sense, CSL has a negligible effect on the reflection probability. The time scale $t_2$ can be understood as a cut off on the time integration in (\ref{EB}). This implies that there is a corresponding momentum spread in the reflected peak of order $\sqrt{Dt}$ - the same as we found for the transmitted peak. This agrees with what we observe in Fig.~\ref{F1}

Finally let us consider the $a$-dependent term in (\ref{KA}). This becomes significant when 
\begin{align}
\frac{\hbar^2}{a^2} \lesssim Dt \sim \Delta p.
\end{align}
If this is the case then the first exponential term in Eq.(\ref{EB}) along with the condition (\ref{momcon}) imply that the reflected component is already heavily suppressed by the barrier being smooth rather than sharp. 

In conclusion the dominant effect of CSL during the process of quantum mechanical refection is momentum diffusion. The magnitude of the momentum diffusion is the same order as that undergone by a free wave packet travelling for the same amount of time $t$ and satisfying CSL dynamics.

%%%%%%%%%%%%%

\section{Two non-interacting identical particles}
\label{S2}
So far we have looked at systems composed of only 1 particle. Now we consider 2 particles. For simplicity we assume that there is no conventional interaction between the 2 particles and examine only the behaviour resulting from spontaneous localization. We take the particles to be identical spinless bosons, each initially in the same Gaussian state centred about $x=0$ with zero expected momentum.

The two particle state is represented by
\begin{align}
|\psi\rangle = \int d{x_1}dx_2\; \psi({x_1},x_2) \frac{1}{\sqrt{2}}\hat{a}^{\dagger}({x_1})\hat{a}^{\dagger}({x_2})|0\rangle,
\end{align}
where $1$ and $2$ label the two particles. Since the two particles are identical their joint wavefunction $\psi(x_1,x_2)$ is symmetric under interchange of coordinates. Improper position states are given by $|x_1,x_2\rangle  = \hat{a}^{\dagger}({x_1})\hat{a}^{\dagger}({x_2})|0\rangle/\sqrt{2}$ and the coordinate space representation of the density matrix is $\langle x_1,x_2|\hat{\rho}|y_1,y_2\rangle$.

Following the same method outlined in Sec.~\ref{S1} we arrive at the two particle master equation
\begin{align}
\frac{\partial}{\partial t}\rho_t =& \frac{i\hbar}{2m}\left(\frac{\partial^2}{\partial x_1^2} - \frac{\partial^2}{\partial y_1^2}\right)\rho_t + \frac{i\hbar}{2m}\left(\frac{\partial^2}{\partial x_2^2} - \frac{\partial^2}{\partial y_2^2}\right)\rho_t \nonumber\\
&-\frac{D}{\hbar^2}\left[(x_1-y_1)^2+(x_1-y_2)^2+(x_2-y_1)^2\right.\nonumber\\
&\quad\quad\quad\left.+(x_2-y_2)^2-(x_1-x_2)^2-(y_1-y_2)^2\right]\rho_t,
\label{M2}
\end{align}
where we have used that $x_i-y_i$ for $i = 1,2$, and $x_1-x_2$ are typically much smaller than $1/\sqrt{\alpha}$. In other words the particles are each sufficiently localized and close together.

The propagator for Eq.(\ref{M2}) is readily shown to be given by
\begin{align}
J(x_1&,y_1,x_2,y_2,t|x_1',y_1',x_2',y_2',t')\nonumber\\
 =& J(x_1,y_1,t|x_1',y_1',t')J(x_2,y_2,t|x_2',y_2',t') \nonumber \\ 
& \times \exp \left\{ -\frac{D(t-t')}{3\hbar^2}\left[ (x_1-y_2)^2+(x_1-y_2)(x_1'-y_2')+(x_1'-y_2')^2 \right] \right\}\nonumber\\
& \times \exp \left\{ -\frac{D(t-t')}{3\hbar^2}\left[ (x_2-y_1)^2+(x_2-y_1)(x_2'-y_1')+(x_2'-y_1')^2 \right] \right\}\nonumber\\
& \times \exp \left\{ +\frac{D(t-t')}{3\hbar^2}\left[ (x_1-x_2)^2+(x_1-x_2)(x_1'-x_2')+(x_1'-x_2')^2 \right] \right\}\nonumber\\
& \times \exp \left\{ +\frac{D(t-t')}{3\hbar^2}\left[ (y_1-y_2)^2\; +(y_1-y_2)(y_1'-y_2')\;+(y_1'-y_2')^2 \right] \right\},
\end{align}
where the result is expressed in terms of the one particle propagator given in (\ref{propx}). This result follows from the one particle result by inspection although it could in principle be derived using path-integral methods (see e.g.~Refs \cite{CALD,ANAS} for use of path-integral methods for evaluation of the density matrix propagator).

The initial wavefunction is taken to be
\begin{align}
\psi(x_1,x_2) =& \frac{1}{\sqrt{2\pi\sigma^2}}\exp\left\{-\frac{1}{4\sigma^2}x_1^2\right\}
\exp\left\{-\frac{1}{4\sigma^2}x_2^2\right\},
\end{align}
from which the initial density matrix follows
\begin{align}
\rho_0(x_1,y_1,x_2,y_2) = \psi(x_1,x_2) \psi^*(y_1,y_2).
\end{align}
We suppose that this state results from trapping the two particles in the same harmonic trap. Once the trapping potential is turned off the particles are left to undergo the free particle dynamics described by the CSL model. The joint probability distribution for subsequently measuring the two particles in positions $x_1$ and $x_2$ is given by
\begin{align}
P_t(x_1,x_2) =& \rho_t(x_1,x_1,x_2,x_2)\nonumber\\
=&\int dx'_1dy'_1dx'_2dy'_2 J(x_1,x_1,x_2,x_2,t|x_1',y_1',x_2',y_2',0)\rho_0(x'_1,y'_1,x'_2,y'_2).
\end{align}
All the Gaussian integrals are straightforward to perform. The result is simplest when expressed in term of the variables
\begin{align}
X = \frac{x_1+x_2}{2} \quad \text{and} \quad \xi = x_1-x_2.
\end{align}
In terms of these variables the probability distribution function factorizes and we find
\begin{align}
P_t(X,\xi) =& \frac{1}{2\pi\sigma^2 L^{1/2}}\exp\left\{ -\frac{1}{\sigma^2L}\left(
\frac{\hbar^2t^2}{4m^2\sigma^4} +1\right)X^2\right\}\nonumber\\
&\times \exp\left\{ -\frac{1}{\sigma^2L}\left(\frac{4Dt^3}{3m^2\sigma^2}+\frac{\hbar^2t^2}{4m^2\sigma^4} +1\right)\left(\frac{\xi}{2}\right)^2\right\},
\label{jointP}
\end{align}
where
\begin{align}
L = \frac{D\hbar^2t^5}{3m^4\sigma^6} + \frac{\hbar^4t^4}{16m^4\sigma^8}
+\frac{4Dt^3}{3m^2\sigma^2} + \frac{\hbar^2 t^2}{2m^2\sigma^4} + 1.
\end{align}

The interesting feature of (\ref{jointP}) is that the spread of the distribution of $X$ is different from the spread of the distribution of $\xi/2$. If the particles were undergoing spontaneous localization independently, each evolving according to Eq.(\ref{M1}) we would find that $X$ and $\xi/2$ have the same distribution at all times. In fact they would each have a Gaussian distribution with variance $\sigma^2K/2$ (see Eq.(\ref{K})).
\begin{figure}[h]
        \begin{center}
        	\includegraphics[width=15cm]{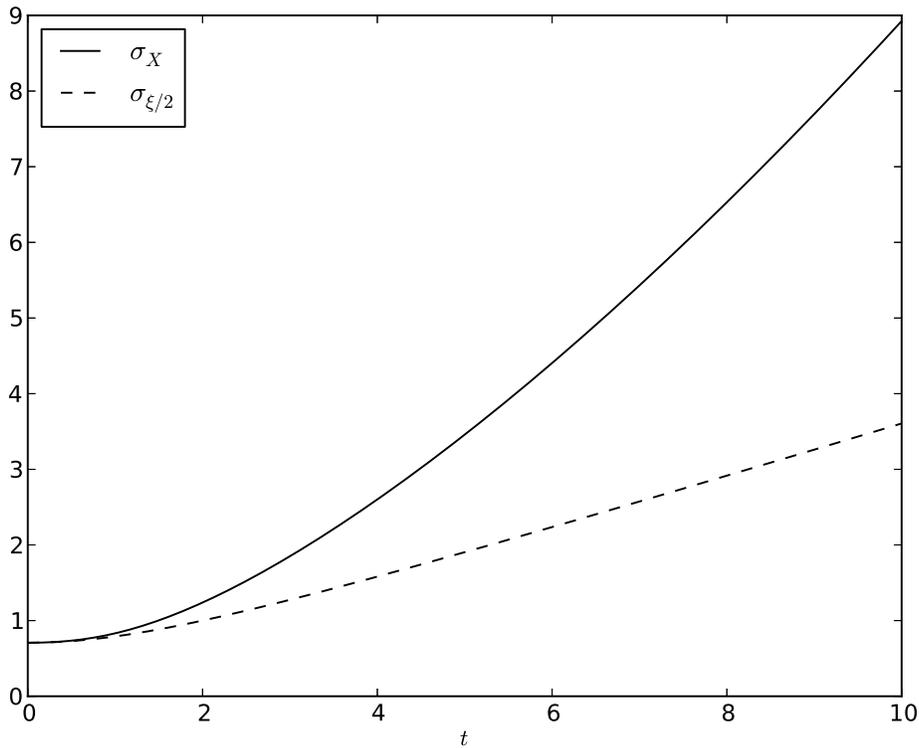}
        \end{center}
\caption{Standard deviation in $X = (x_1 + x_2)/2$ and $\xi/2 = (x_1 - x_2)/2$ with time, for two particles released from a harmonic trap.}
\label{F3}
\end{figure}

The behaviour of the standard deviations in $X$ and $\xi/2$ with time is shown in Fig.~(\ref{F3}). Units are chosen such that $\hbar = m = \sigma = 1$ and the parameter $D$ is set to $0.1$. The initial standard deviation is $1/\sqrt{2}$ in these units. This grows more rapidly with time for $X$ than for $\xi/2$. We can read off the long time behaviour of the standard deviation from (\ref{jointP}). We find that $\sigma_{\xi/2}\propto t$ and $\sigma_{X}\propto t^{3/2}$.

If the interactions between particles can be kept to a minimum (or quantified precisely) we can imagine that this effect could be measured by releasing the particles from the ground state in a harmonic trap then allowing the particles to move freely in 1 dimension before measuring their positions on a device/screen below the trap.

To get an idea of the numbers involved consider two particles each of $10^8$ nucleon masses in a trap of width $10^{-8}{\rm m}$. The particles are released and left for $10{\rm s}$ before having their positions measured. Here we find that the standard deviation in $X$ is of order $10\%$ larger than the standard deviation in $\xi/2$ using the GRW parameters. 

The particles do not need to be identical. Provided that the operator $\hat{N}(x)$ in Eq.(\ref{CSL}) is replaced by the total matter density operator
\begin{align}
\hat{M}(x) = \sum_k \frac{m_k}{m_0} \hat{N}_k(x),
\end{align}
where $k$ denotes different particle species, the result is the same. (Equation (\ref{jointP}) will only apply when the 2 particles have the same mass $m$.)

The physical reason for this effect is that the diffusions undergone by each particle are correlated by the way the localization mechanism works on the total number density state rather than individually on each particle. This prevents the particles from spreading too far apart even though the system as a whole will diffuse.

We also note that tracing out one of the two particles in (\ref{jointP}) results in the remaining particle behaving precisely as described by the one particle propagator (\ref{propx}) .

%%%%%%%%%%%%%

\section{Discussion}
\label{SDISC}
We have demonstrated the use of the density matrix propagator as a way of solving the continuous spontaneous localization model in a range of experimental situations in the case where the localization length scale can be regarded as large. The technique involves constructing solutions by evolving an initial density matrix according to Eq.(\ref{rhosol1}) and its generalizations to include more particles. Interactions, such as those with a classical potential can be added perturbatively in a standard way. 

The first situation that we considered was the two slit experiment. We demonstrated that the interference pattern predicted by standard quantum theory becomes gradually less visible as we either increase the mass of the particle, increase the slit separation, or increase the slit widths. In particular, increasing the mass has the joint effect of increasing the rate of localization through Eq.(\ref{lamb}) and increasing the time taken for the two wave packets to reach a state of overlap (\ref{talb}).

The second example made use of perturbation theory to examine the process of scattering of a particle from a classical potential barrier. The potential height was assumed to be small and treated perturbatively to second order. The result was that there was no significant change in the probability of reflection. The dominant effect of SL was to cause momentum diffusion. The amount of diffusion for the reflected peak was of the same order of magnitude as that of the transmitted peak which is dominated by a zeroth order contribution. 

The final example was to consider the CSL evolution of 2 non-interacting particles. To give a concrete example we chose the 2 particles to be initially located in the same trap with the same initial wave function. Once the trap is switched off the 2 particles evolve freely. However, their behaviour is coupled by the CSL dynamics and this has the effect of making it more likely that the particles will be subsequently measured closer together than would be expected if the particles dispersed independently. We quantified this in the form of a joint probability distribution for the positions of the two particles.

In general we have demonstrated the use of the density matrix propagator as a powerful tool for solving the CSL model. The results derived are, in principle, experimentally observable and offer tests of CSL against standard quantum theory.

\section*{Acknowledgements}
I would like to thank Jonathan Halliwell for helpful discussions and comments. This work was supported by EPSRC Research Grant No.~EP/J008060/1.

\end{document}